\documentclass[acmtog,screen,balance=true]{acmart}
\setcopyright{acmlicensed}
\usepackage{amsmath,bm}
\usepackage{mathtools}
\usepackage{xspace}
\usepackage{tikz}
\usetikzlibrary{spy}
\usepackage{microtype}
\usepackage[percent]{overpic}
\usepackage[outline]{contour}
\usepackage{tabularx}
\usepackage{multirow}
\usepackage{slantsc}
\contourlength{0.1em}%
\usepackage{tcolorbox}
\newcommand{\IGNORE}[1]{}

\newcommand{\REMOVE}[1]{} % for final version
\newcommand{\ADD}[1]{#1} % for final version
\def\equationautorefname~#1\null{%
  Eq.~(#1)\null
}

\newcommand{\Paragraph}[1]{\subsubsection*{#1}}

\newcommand{\Point}{\mathbf{x}}
\newcommand{\Dir}{\mathbf{\omega}}
\newcommand{\DirE}{\mathbf{\omega}^e}

\newcommand{\Radiance}{L}

\newcommand{\RadianceAt}[2]{\Radiance(#1,#2)}
\newcommand{\Transmittance}[1]{\tau(#1)}

\newcommand{\CoeffScatter}{\sigma_s}

\newcommand{\CoeffExtinct}{\sigma_t}

\newcommand{\ScatterFunction}[1]{f_\rho(#1)}
\newcommand{\PhaseFunction}[2]{\rho(#1, #2)}

\newcommand{\dif}{\mathrm{d}}

\newcommand{\Est}[1]{E[#1]}
\newcommand{\RandV}[1]{\xi(#1)}

\newcommand{\Pdf}[1]{p(#1)}
\newcommand{\PdfDist}[1]{p_t(#1)}
\newcommand{\PdfDir}[1]{p_w(#1)}
\newcommand{\PdfM}[1]{p_m(#1)}
\newcommand{\PdfE}[1]{p_e(#1)}
\newcommand{\PdfP}[1]{p_{\rho}(#1)}

\newcommand{\Emission}{\Radiance_e}
\newcommand{\VolumeEmission}{\Radiance_{v}}

\newcommand{\SPos}{\mathbf{x}}
\newcommand{\SPath}{\ensuremath{\overline{\mathbf{z}}}\xspace}
\newcommand{\Prior}[1]{\ensuremath{\hat{#1}}\xspace}

\newcommand{\PPathSpace}{\Omega}

\newcommand{\PContri}{f}

\newcommand{\AiMatrix}{\mathbf{A}^{+}}
\newcommand{\AdMatrix}{\mathbf{A}^{o}}
\newcommand{\PMatrix}{\mathbf{P}}

\newcommand{\AngleSpace}{S}

\newcommand{\PathGraph}{G}
\newcommand{\VertexUnion}{V}
\newcommand{\EdgeUnion}{E}
\newcommand{\LPos}{\mathbf{y}}
\newcommand{\Edge}{\mathbf{e}}
\newcommand{\CEdge}{\Edge_C}
\newcommand{\LEdge}{\Edge_L}
\newcommand{\NEdge}{\Edge_N}
\newcommand{\CEdgeUnion}{\EdgeUnion_C}
\newcommand{\LEdgeUnion}{\EdgeUnion_L}
\newcommand{\NEdgeUnion}{\EdgeUnion_N}
\newcommand{\PixelValue}{I}
\newcommand{\LPosUnion}{\VertexUnion_y}
\newcommand{\SPosUnion}{\VertexUnion_x}
\newcommand{\RadianceOut}{\ensuremath{\overline{\Radiance}}\xspace}
\newcommand{\RadianceOutVec}{\ensuremath{\overline{\mathbf{\Radiance}}}\xspace}
\newcommand{\Indirect}{\Radiance^\mathbf{+}}
\newcommand{\IndirectIn}{\Indirect}
\newcommand{\IndirectInVec}{\mathbf{\Indirect}}
\newcommand{\IndirectOut}{\ensuremath{\overline{\Indirect}}\xspace}
\newcommand{\IndirectOutVec}{\ensuremath{\overline{\mathbf{\Indirect}}}\xspace}
\newcommand{\DirectEmission}{\Radiance^{\mathbf{o}}}
\newcommand{\DirectEmissionIn}{\DirectEmission}
\newcommand{\DirectEmissionInVec}{\mathbf{\DirectEmissionIn}}
\newcommand{\DirectEmissionOut}{\ensuremath{\overline{\DirectEmission}}\xspace}
\newcommand{\DirectEmissionOutVec}{\ensuremath{\overline{\mathbf{\DirectEmission}}}\xspace}

\newcommand{\MISweight}{w}
\newcommand{\cluster}{C}

\newcommand{\bunnycloud}{\textsc{Bunny Cloud}}

\newcommand{\disneycloud}{\textsc{Disney Cloud}}

\newcommand{\mtscolorsmoke}{\textsc{Colored Smoke}}
\newcommand{\industrysmoke}{\textsc{Industry Smoke}}
\newcommand{\dustshockwave}{\textsc{Dust Shockwave}}

\newcommand{\buddha}{\textsc{Buddha}}
\newcommand{\trafficlight}{\textsc{Traffic Light}}
\newcommand{\foggyforest}{\textsc{Foggy Forest}}
\newcommand{\goldengate}{\textsc{Golden Gate}}
\newcommand{\MaxDepth}{D}

\usepackage{pbox}
\usepackage{booktabs}
\newlength{\mygrid}%
\newlength{\myimg}%

\begin{document}

\title[Rendering Participating Media Using Path Graphs]{Rendering Participating Media Using Path Graphs}

%% Authors
\author{Becky Hu}
% \email{bh456@cornell.edu}
\affiliation{%
  \institution{Cornell University}
  \country{USA}
}

\author{Xi Deng}
\affiliation{%
  \institution{Cornell University}
  \country{USA}
}

\author{Fujun Luan}
\affiliation{%
  \institution{Adobe Research}
  \country{USA}
}

\author{Miloš Hašan}
\affiliation{%
  \institution{Adobe Research}
  \country{USA}
}

\author{Steve Marschner}
\affiliation{%
  \institution{Cornell University}
  \country{USA}
}

\renewcommand{\shortauthors}{Firstauthor et al.}

\begin{abstract}
Rendering volumetric scattering media, including clouds, fog, smoke, and other complex materials, is crucial for 
realism in computer graphics. Traditional path tracing, while unbiased, requires many long path samples to converge in scenes with scattering media,
and a lot of work is wasted by paths that make a negligible contribution to the image. Methods to make better use of the information learned during path tracing range from photon mapping to radiance caching, but struggle to support the full range of heterogeneous scattering media.
This paper introduces a new volumetric rendering algorithm that extends and adapts the previous \emph{path graph} surface rendering algorithm. Our method leverages the information collected through multiple-scattering transport paths to compute lower-noise estimates, increasing computational efficiency by reducing the required sample count. Our key contributions include an extended path graph for participating media and new aggregation and propagation operators for efficient path reuse in volumes.  Compared to previous methods, our approach significantly boosts convergence in scenes with challenging volumetric light transport, including heterogeneous media with high scattering albedos and dense, forward-scattering translucent
materials, under complex lighting conditions. 
\end{abstract}

%%
%% Generate your CCSCML using http://dl.acm.org/ccs.cfm.
%%

%%%%%%%%%%%%%%%%%%%%%%%%%%%%%%%%%%%%%%%%%%%%%%%%%%%%%%%%%%%%
% ACM categories, keywords, etc.
%%%%%%%%%%%%%%%%%%%%%%%%%%%%%%%%%%%%%%%%%%%%%%%%%%%%%%%%%%%%

    

%% Teaser figure that appears on the top of the article.
%% Uncomment the includegraphics line to include an actual teaser image.
%% Make sure to fill out a description for accessibility
\begin{teaserfigure}
  \includegraphics[width=\textwidth]{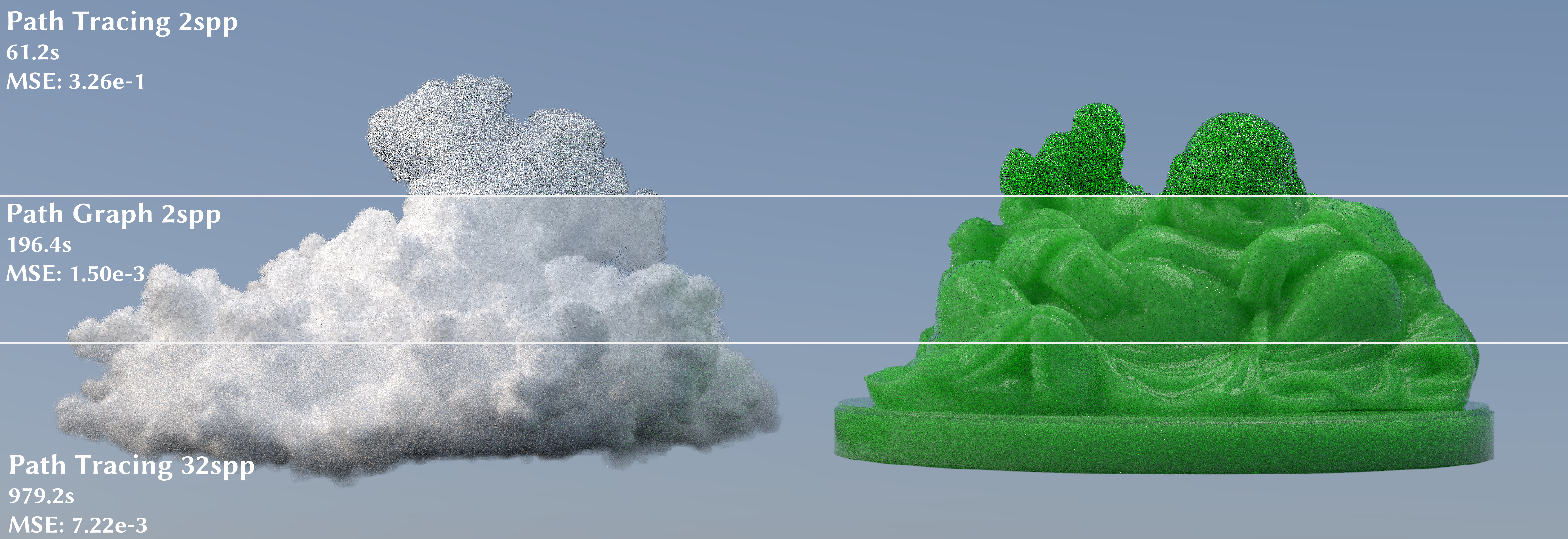}
  \caption{A heterogeneous cloud, left, and a homogeneous buddha statue with a dielectric interface, right, both illuminated by a sun-sky model and an area light. At the top we show an equal-sample comparison between path tracing and our method (path tracing + path graph); our method reduces variance significantly. At the bottom we show that using our method with a very small number of samples can both take less time and produce a more accurate result than using a larger number of samples with path tracing.}
  \Description{This is the teaser figure for the article.}
  \label{fig:teaser}
\end{teaserfigure}

\maketitle

%% The actual document with your content starts here
\section{Introduction}
Volumetric scattering media 
(such as clouds, fog, smoke, juice, soap, or marble)
are common in real-world scenes, and accurate rendering of such materials is key to realism in scenes with scattering atmospheres such as clouds, fog, or smoke; turbid fluids like juice, milk, or seawater; or translucent solids like marble. However, simulating light transport in participating media presents well-known challenges: compared to scenes with only surfaces there is an increased number of scattering events, especially in media with anisotropic phase functions, and it is challenging to sample fast-varying optical properties, especially when optical density varies spectrally.

Monte Carlo path tracing is widely used for rendering scenes with participating media; while unbiased, it requires a large number of samples to converge. High sample counts become particularly problematic when artists need fast turnaround during the scene design process. 
%Recent neural d
Denoisers improve convergence to some extent; however, better sampling methods with lower variance are desirable independent of whether denoising is used. The high-level goal of this paper is to extract as much information as possible from a very small number of volumetric scattering paths so that we can obtain useful images quickly and ultimately achieve faster convergence.

%The light that travels through a 
Participating media are characterized by long multiple-scattering transport paths. By tracing 
these paths
%such a long path 
through the volume, 
a path tracer collects a lot of valuable information about the scene, but then computes a single pixel contribution and discards all the other information. %collected while tracing it. 
While this has the benefit of requiring minimal storage, it clearly feels suboptimal in terms of computational efficiency, and many methods have been proposed to make better use of the traced paths. This path information can be stored in the scene and then gathered
%to create smoother estimations 
by density estimators, including photon points \cite{Jensen:1995:Photon, Jensen:1996:Photon}, photon beams \cite{Jarosz:2008:Beam}, and photon primitives \cite{Bitterli:2017:Points, Deng:2019:PhotonSurface}; weighted by multiple importance sampling \cite{Veach:1995:Optimally} to combine different sampling techniques \cite{Georgiev:2012:Importance, Hachisuka:2012:Path, Krivanek:2014:Unifying}; or cached to guide smarter future sampling \cite{Herholz:2019:VPathGuiding, Bitterli:2020:ReSTIR, Lin:2021:VolumeReSTIR}. However, the reuse of information in these methods is constrained to be relatively local, among spatially close paths or paths from neighboring pixels. We would prefer more global information sharing, especially in scenes with complex, long paths where the information is hidden behind multiple bounces 
%with spatially varying optical properties, 
or in scenes dominated by indirect illumination where the majority of the paths fail to find the light source.
In fact, such cases are ubiquitous in volumetric scenes, such as subsurface-scattering materials with dielectric boundaries, or clouds and smoke with challenging lighting conditions and high single-scattering albedos. 

Recent work on path graphs \cite{Deng:2021:PathGraph} has shown the benefits of sharing information globally across all traced paths in scenes dominated by indirect illumination.  
As illustrated in \autoref{fig:structure}, the path graph framework records the information of all sampled paths during the path sampling process and builds a  graph over these paths. Iterative refinement (through aggregation and propagation) is then applied to the path graph to refine the radiance returned from each path. 
This approach is beneficial because it not only shares radiance among spatially nearby shading points, but also iteratively propagates the updated radiance along paths to refine the estimates at distant shading points, making extensive use of the initial sampled paths. However, this technique is limited to surfaces.
%Therefore, it is independent of sampling techniques and the computation on the path graph does not require access to the scene geometry nor the tracing of additional rays.

Our goal is to develop a new method based on the path graph framework that applies to volumetric scattering. This requires several changes to the framework, since the original path graph method relies on surface shading points and related concepts such as normals, BRDFs, and their importance sampling pdfs. Furthermore, light extinction (transmittance) is unique to volumes and introduces important subtleties when extinction varies spectrally.
%not among the concepts supported by the original method. 
%However, w
We show how the framework can be extended to volumetric rendering and demonstrate that the resulting method significantly improves convergence across a variety of scenes with challenging volumetric transport, including heterogeneous media with high scattering albedos, subsurface scattering materials, and challenging lighting conditions. %The main contributions of our work include
% \XD{need to refine this list}
% \begin{itemize}
%     \item Expanding the application of the path graph framework for image reconstruction from sampled paths to include scenarios involving participating medium.
%     \item Derive the aggregation and propagation operators of the path graph for volumetric rendering.
%     \item \XD{Any acceleration technique?}
%     \item \XD{Maybe we can try to demonstrate a example of combining volume path graph with surface path graph if time allowed}
%     % \item faster clustering for volume \XD{not sure}
%     \item We implement a practical volumetric path graph algorithm for robust light transport in participating media.
% \end{itemize}

% This inspired us to utilize the concept of a path graph, which allows for global information sharing through long paths by reusing irradiance of neighboring shading points and propagating the updated radiance along the path. In this work, we develop methods that enable global sharing in volumetric rendering and explore the benefits gained from such sharing.

% In addition to paths reuse idea, smart sampling technique, which navigate the scattering direction proportion to steady state irradiance, like path guiding ~\cite{Herholz:2019:VPathGuiding} has also been used to reduce the noise. 

\section{Related Work and Background}

\subsection{Volumetric rendering}
Volumetric rendering is typically accomplished by solving the radiative transfer equation (RTE) ~\cite{Kajiya:1989:Rendering} using Monte Carlo estimation. Naive path tracing is the most popular rendering algorithm for participating media due to its unbiased nature and applicability to a wide variety of volumes. However, naive path tracing suffers from a low convergence rate in challenging lighting conditions, anisotropic scattering properties, and heterogeneous optical properties.

To address the low convergence rate issue, path samples can be reused. Bidirectional path tracing \cite{Veach:1994:Bidirectional, Veach:1997:Robust, Lafortune:1993:Bidirectional} (BPT) samples paths from both the camera and light sources, then connects the sub-paths and combines them using multiple importance sampling \cite{Veach:1995:Optimally}. Paths can be also reused to improve sampling techniques. Volumetric path guiding \cite{Herholz:2019:VPathGuiding} caches paths to estimate adjacent transport solutions and uses them to navigate future sampling. The idea of spatiotemporal reservoir resampling \cite{Bitterli:2020:ReSTIR} has also been applied in fast volume rendering  \cite{ Lin:2021:VolumeReSTIR}.

Methods that reuse path samples by connecting paths sampled from sensors and from sources using photon density estimators, such as photon mapping \cite{Jensen:1996:Photon, Jensen:1998:Efficient}, are more efficient, although they introduce bias from blurring kernels. To mitigate this bias, researchers have proposed techniques like photon beams, sensor beams \cite{Jarosz:2008:Beam, Jarosz:2008:Beama}. Similarly to BPT, research has sought to combine these density estimators with each other and with unbiased techniques \cite{Georgiev:2012:Light, Hachisuka:2012:Path, Krivanek:2014:Unifying}. Over time, unbiased photon density estimators like photon planes, photon volumes \cite{Bitterli:2017:Points}, and even photon surfaces \cite{Deng:2019:PhotonSurface} have been developed. However, the effectiveness of these unbiased density estimators is limited to homogeneous volumes.
%, and transforming them to heterogeneous volumes in practice is non-trivial.

Neural networks have been used for fast prediction in participating media.  For instance Hu et al.~\cite{Hu:2023:DeepVolume} applied radiance-predicting neural networks to store directional light transmittance. However, this approach is limited to specific lighting conditions.

\subsection{Path Reuse Techniques}
Beyond volumetric rendering, many path reuse technique has been applied in surface rendering scenario. Early work in irradiance caching \cite{Ward:1988:Ray} stores the irradiance estimates for future interpolation of efficient indirect illumination. \cite{Krivanek:2008:Practical} extends to cache radiance allowing indirect lighting computation with the presence of glossy surface. \cite{Keller:2014:Path}
directly computes weighted average of from noisy radiance estimates on the path in Monte
Carlo path tracing. \cite{West:2020:ContinuousMIS} later extends the work by applying continuous multiple important sampling in combining the noise irradiance estimates, \cite{Deng:2021:PathGraph} further improved the efficiency of filtering using clusters for aggregation and iteratively propagate the updated refinement from indirect bounce to pixel. 

\section{Method}
In this section, we will briefly review the key ideas and concepts of the path graph framework (\autoref{sec:key-idea}), and derive our path graph operators for participating media starting from standard radiative transfer equations (\autoref{sec:rte-medium}-\autoref{sec:pg-operators}).

\subsection{Path Graph Framework}
\label{sec:key-idea}
\begin{figure}
    \centering
\includegraphics[page=2,width=0.48\textwidth]{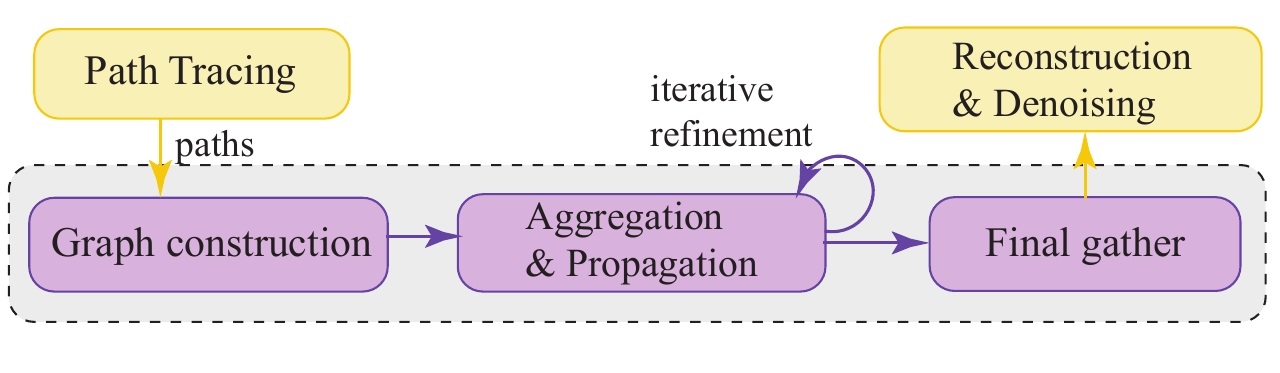}
    \caption{Our method fits  between the traditional steps of path tracing and denoising. This means that it can provide additional benefits on top of (rather than replacing) techniques like 
    %modern Monte Carlo 
    neural
    denoisers or advanced path sampling. Unlike the surface Path Graph, the final gather is optional in the volume case. The final gather serves to remove correlations between neighboring pixels, which are very subtle in volumes since path vertices are scattered in the volume.}
    \label{fig:structure}
\end{figure}

In the rendering framework of path graphs (\autoref{fig:structure}), the path tracing phase is instrumented to record enough intermediate results to recompute the pixel value given a change to any outgoing radiance value along the path. The key idea is to update these intermediate results in an iterative process where nearby shading points exchange information and then recompute the radiance values from the updated values. We think of these intermediate values as variables attached to the vertices and edges of a graph in which every shading point is a vertex and the edges record all the relevant relationships between vertices including which values are used to update which other values and which points are “nearby” so that they exchange information.

%\STEVE{It seems natural to me to move the text from S 3.5 that defines the path graph up to this section.}

In the following context, we refer to a path graph $\PathGraph = \langle \VertexUnion, \EdgeUnion \rangle$  as an information sharing data structure built upon a collection of complete light paths which are sampled by path tracing of a full resolution image at one sample per pixel. $\VertexUnion$ is the union of all the light points $\LPosUnion$ and shading points $\SPosUnion$.  The vertices are connected to each other by three different types of edges in $\EdgeUnion$:  continuation edges $\CEdgeUnion$, light edges $\LEdgeUnion$ and neighbor edges $\NEdgeUnion$. A \emph{continuation edge} $\CEdge \in \CEdgeUnion$ stands for the propagation of a radiance value from one shading point to another; the outgoing value on a point becomes the incoming value on another. A \emph{light edge} $\LEdge \in \LEdgeUnion$ connects a light point with a shading point.  It propagates emitted light from source points to surface points. A \emph{neighbor edge} $\NEdge \in \NEdgeUnion$ connects shading points with other shading points in their spatial neighborhood (cluster).

The construction of light edges and continuation edges is done simultaneously with path tracing. The neighbor edges can be constructed by a simple nearest neighbor clustering of all shading points. Given $N$ shading points and a desired number of shading points per a cluster, $K$, approximately $N/K$ shading points are chosen as cluster centers, and all other shading points are assigned to the clusters with the nearest cluster centers. This process is repeated for large clusters until the number of shading points per cluster is relatively even.

Shading points within a cluster are assumed to share the same incoming radiance distribution, and the neighbor edges serve as bridges for aggregation across those shading points. We treat the neighbors’ incoming radiance samples as samples of the point’s incoming radiance. Therefore, intuitively, we can treat each of the $N$ points in the cluster as a sampling technique with its own pdf, and combine the $N$ techniques using multiple importance sampling. The resulting improved estimates can further be used to improve estimates in other clusters that depend on them, until convergence is reached. In the following sections, we turn these intuitions into a precise estimator for volume scattering media.

 %\cite{Deng:2021:PathGraph} have shown that clustering shading points into disjoint clusters is essential in order to keep the aggregation computation efficient.

\subsection{Radiative Transfer Equation in a Medium}
\label{sec:rte-medium}

We will start with the radiative transfer equation (RTE) in a medium (\autoref{sec:rte-medium}) and review the Monte Carlo solution to the RTE (\autoref{sec:monte-carlo-rte}). We then derive the path graph propagation and aggregation operators that apply to participating media. 
% \STEVE{Would be nice to have a citation for the details of the RTE; then we can be brief in introducing it.}

The radiance propagating in direction $\Dir$ at a point $\Point$
in the medium
can be written as 
%in the form
 \begin{align}
 &\RadianceAt{\SPos}{\Dir} = \Transmittance{\SPos,\LPos}\Emission(\LPos,\Dir) \;+ \nonumber\\
 &\int
 %_\SPos^\LPos 
 \Transmittance{\SPos,\SPos'}\left(\VolumeEmission(\SPos', \Dir) + \int_{S^2}  \ScatterFunction{\SPos',\Dir,\Dir'}\RadianceAt{\SPos'}{\Dir'} \dif \Dir' \right) \dif \SPos', \label{eq:standard-RTE}
 \end{align}
 %\BH{should the term before the addition sign be $\Transmittance{\SPos,\LPos}\Emission(\LPos,-\Dir)$ (so negating the direction in the emitted radiance term)?}
 where $\ScatterFunction{\SPos',\Dir,\Dir'} = \CoeffScatter(\SPos')\PhaseFunction{\SPos'}{\Dir,\Dir'}$, 
$\LPos$ is the visible surface point in the direction $-\Dir$ from $\SPos$ and the outer integral is along the line segment joining $\SPos$ and $\LPos$.
$\Emission(\LPos, \Dir)$ 
%stands for 
is
the surface emitted radiance 
%along direction $\Dir$ from a point
at
$\LPos$, $\VolumeEmission(\SPos', \Dir)$ is the volume emitted radiance per unit length at $\SPos'$, $\PhaseFunction{\SPos',\Dir}{\Dir'}$ is the phase function of the medium at $\SPos'$, 
%which gives the rate at which energy is scattered into direction $\Dir$ from $\Dir'$, 
$\Transmittance{\SPos,\LPos} = e^{-\int_{\SPos}^{\LPos} \CoeffExtinct(\Point') \dif \SPos'}$ 
%characterizes the loss of energy when the pencil of radiance traverses a thickness of $||\Point-\Point'||$ along its direction of propagation, 
is the transmittance between two points, and $\CoeffExtinct(\SPos)$ and $\CoeffScatter(\SPos)$ are the attenuation and scattering coefficients of the medium.
%optical properties of the medium which characterize the energy loss rate of a beam when propagating through a unit thickness of medium.
 % By putting the direct radiance and emission together we rearrange the \autoref{eq:standard-RTE} as
 % \begin{align}
 %      &\RadianceAt{\SPos}{\Dir} = \Transmittance{\SPos,\LPos} \Emission(\LPos,\Dir) + \int \Transmittance{\SPos,\SPos'}\Emission(\SPos')  \dif \SPos'+  \\
 % &\int  \Transmittance{\SPos,\SPos'}\CoeffScatter(\SPos')\int \PhaseFunction{\SPos'}{\Dir,\Dir'} \RadianceAt{\SPos'}{\Dir'} \dif \Dir' \dif \SPos' \label{eq:standard-RTE}
 % \end{align}
 
For compactness of notation we encapsulate the two nested integrals in \autoref{eq:standard-RTE} inside linear operators $\mathcal{K}$ and $\mathcal{G}$:
\begin{align}
\left[\mathcal{K}f\right](\SPos, \Dir) &= \int_{S^2}  \ScatterFunction{\SPos,\Dir,\Dir'}f(\SPos,\Dir') \dif \Dir'
 \label{eq:scattering-op}\\
\left[\mathcal{G}f\right](\SPos, \Dir) &= \int \Transmittance{\SPos,\SPos'}f(\SPos, \Dir) \dif \SPos'.
 \label{eq:propagation-op}
\end{align}
so that \autoref{eq:standard-RTE} reads
\begin{align}
\Radiance &= Q + \mathcal{G}(\VolumeEmission + \mathcal{K}\Radiance).
\label{eq:operator-RTE}
\end{align}
where $Q(\SPos,\Dir) = \Transmittance{\SPos,\LPos}\Emission(\LPos,\Dir)$. This operator notation is inspired by \citet{Arvo:1993:Linear}.

To align our transport equations with the quantities stored in the path graph,  
 %Before we go over the details of solving the standard radiative transfer equation in \autoref{eq:standard-RTE}, we will introduce our operator form of \autoref{eq:standard-RTE} in the following context and will expand it by one step to get a form of the radiative transfer equation with terms corresponding to our path graph operators.
we rearrange the RTE  in two ways.  First, we introduce scattered radiance $\RadianceOut$ and write \autoref{eq:operator-RTE} in two steps as:
\begin{equation}
\left\{\quad
\begin{aligned}
 \RadianceOut &= \mathcal{K}\Radiance\\
 \Radiance &= Q + \mathcal{G}(\VolumeEmission + \RadianceOut).
\end{aligned}
\right.
\end{equation}
Like volume emission, $\RadianceOut$ is a radiance per unit length with units $W/(m^3 \cdot sr)$, whereas $\Radiance$ has units $W/(m^2 \cdot sr)$.  

Second, we write the radiance $\Radiance$ as a sum of $\DirectEmission = Q + \mathcal{G}\VolumeEmission$, which is incoming radiance due to direct illumination and volume emission, and $\IndirectIn = \mathcal{G}\RadianceOut$, the radiance due to multiple scattering.  We also introduce corresponding scattered quantities $\DirectEmissionOut = \mathcal{K}\DirectEmissionIn$ and $\IndirectOut = \mathcal{K}\IndirectIn$.  The RTE then reads
\begin{equation}
\left\{\quad
\begin{aligned}
\IndirectOut &= \mathcal{K}\IndirectIn\\
     \IndirectIn &= \mathcal{G}\DirectEmissionOut + \mathcal{G}\IndirectOut
\end{aligned}
\right.
\end{equation}
In this pair of equations $\DirectEmissionOut$ is known (it is an integral of known quantities) and $\IndirectIn$ is the unknown to solve for, after which the complete solution is simply $\Radiance = \DirectEmissionIn + \IndirectIn$.

The steps of the path graph algorithm will correspond to these operators; \autoref{eq:scattering-op} corresponds to the aggregation step and \autoref{eq:propagation-op} corresponds to the propagation step.  One aggregation step is used to compute $\DirectEmissionOut$; a series of alternating aggregations and propagations serve to solve for $\IndirectIn$; and a final addition computes $\Radiance$ for the pixels of the image.
%\autoref{eq:integral-direct} and \autoref{eq:integral-indirect} are unbiased version of aggregation operators in the path graph \cite{Deng:2021:PathGraph} and    \autoref{eq:integral-RTE_3} is the propagation operator; their  propagation operator in the surface only scene can be think as a sepcial form of \autoref{eq:integral-RTE_3} with a delta integrand. 
 % \STEVE{I just made this up; someone better check that it actually matches the algorithm...} \BH{This is pretty accurate.}

In the next subsection, we will review the general form of Monte Carlo estimators for $\mathcal{K}$ and $\mathcal{G}$, then show
%to the unbiased aggregation operator \autoref{eq:integral-direct},\autoref{eq:integral-indirect} and the propagation  operator \autoref{eq:integral-RTE_3}, then we show 
how the path graph algorithm efficiently approximates these operators from a limited number of sampled paths.

\subsection{Monte Carlo Estimator for RTE}
\label{sec:monte-carlo-rte}
Consider a general integral $\PixelValue = \int_{\PPathSpace} \PContri(z) \dif z$ on the space $\PPathSpace$. Let $z_1,z_2\cdots z_N$ be random sample on $\PPathSpace$ drawn from a probability density function $\Pdf{z}$, and assume $\Pdf{z}$ is nonzero whenever $\PContri(z)$ is nonzero. One can prove that $\RandV{z} = \frac{\PContri(z)}{\Pdf{z}}$ is an unbiased estimator of $\PixelValue$, because
\begin{align}
    \Est{\RandV{z}} =
    \int_{\PPathSpace} \frac{\PContri(z)}{\Pdf{z}} \Pdf{z} \dif z =
    \int_{\PPathSpace} \PContri(z) \dif z = I.\label{eq:mc}
\end{align}
Applying the Monte Carlo estimator \autoref{eq:mc} to the operator $\mathcal{K}$ defined in \autoref{sec:rte-medium}, we find that the scattered quantities $\DirectEmissionOut$ and $\IndirectOut$ can be estimated by a sample drawn from the solid angle space with probability density $\PdfDir{\Dir}$, ideally equal to the phase function (which causes cancellation):
\begin{align}
    \DirectEmissionOut(\SPos', \Dir) \approx \frac{\ScatterFunction{\SPos',\,\Dir,\,\Dir'}\DirectEmissionIn(\SPos', \Dir')}{\PdfDir{\Dir'}},\label{eq:mc-direct} \\
    \IndirectOut(\SPos', \Dir) \approx \frac{\ScatterFunction{\SPos',\,\Dir,\,\Dir'}\IndirectIn(\SPos', \Dir')}{\PdfDir{\Dir'}}.\label{eq:mc-indirect}
\end{align}
The right sides are unbiased estimators, not just approximations. Similarly, to apply the Monte Carlo estimator to the operator $\mathcal{G}$, we take a sample $\SPos'$ along the direction $\Dir$ from $\SPos$:
\begin{align}
    \IndirectIn(\SPos, \Dir) \approx  \frac{\Transmittance{\SPos,\SPos'}\left(\IndirectOut(\SPos',\Dir) + \DirectEmissionOut(\SPos',\Dir)\right)}{\PdfDist{\SPos '|\SPos}},\label{eq:mc-pa}
\end{align}
where again the right side is an unbiased estimator, $\PdfDist{\SPos'|\SPos}$ is the probability density function of sampling distance $||\SPos-\SPos'||$ from $\SPos$, ideally proportional to $\Transmittance{\SPos,\SPos'}$.
The Monte Carlo estimators turn the continuous operators from \autoref{sec:rte-medium} to discrete samples from continuous distributions. In the following section, we will briefly review volumetric path tracing, which is nothing but recursive application of the above estimators, and introduce the construction of a path graph from the resulting path samples.
% \TODO{Need add another paragraph here to summarize the difference in vol path tracing}
% Similar to the aggregation in the surface version of path graphs, the aggregation over the clusters of  the volume path graph also refines the estimation of outgoing direct radiance $\DirectEmissionOut(\SPos_i, \Dir_i)$ and the outgoing indirect radiance $\IndirectOut(\SPos_i,\Dir_i)$, by sharing all its neighbors' incoming samples $\DirectEmissionIn(\SPos_j, \Dir_j)$ and $\IndirectIn(\SPos_j, \Dir_j)$, evaluating their contribution with the scattering property and phase function at $\SPos_i$, and taking a weighted average. In the next subsection, we will talk about the weighting strategy we use for combining the contributions from $\DirectEmissionIn(\SPos_j,\Dir_j)$ and $\IndirectIn(\SPos_j, \Dir_j)$ and write out the formulas for our aggregation and propagation operators.

% Unlike the simple copying operation used in the surface case \autoref{eq:surface-propagation}, propagation for the volume path graph should consider the transmittance term and the probability of sampling $\SPos_{i}$ from $\SPos_{i+1}$ in \autoref{eq:integral-RTE_3}, where
% \begin{align}
%     \IndirectIn(\SPos_{i-1}, \Dir_{i-1}) = \frac{\Transmittance{\SPos_i-\SPos_{i}}}{\Pdf{\SPos_i|\SPos_{i-1}}} \IndirectOut(\SPos_i, \Dir_i)
% \end{align}
% \TODO{Maybe I need to use $\SPos_i$ for $\SPos'$, and $\SPos_{i-1}$ for $\SPos$ in \autoref{eq:integral-RTE_3}}

% In the following subsection, we show how this is applied in estimating pixel value $\PixelValue$.

\subsection{Volume Path Tracing}
\label{sec:vpt}
\begin{figure}
    \centering
\includegraphics[page=1,width=0.475\textwidth]{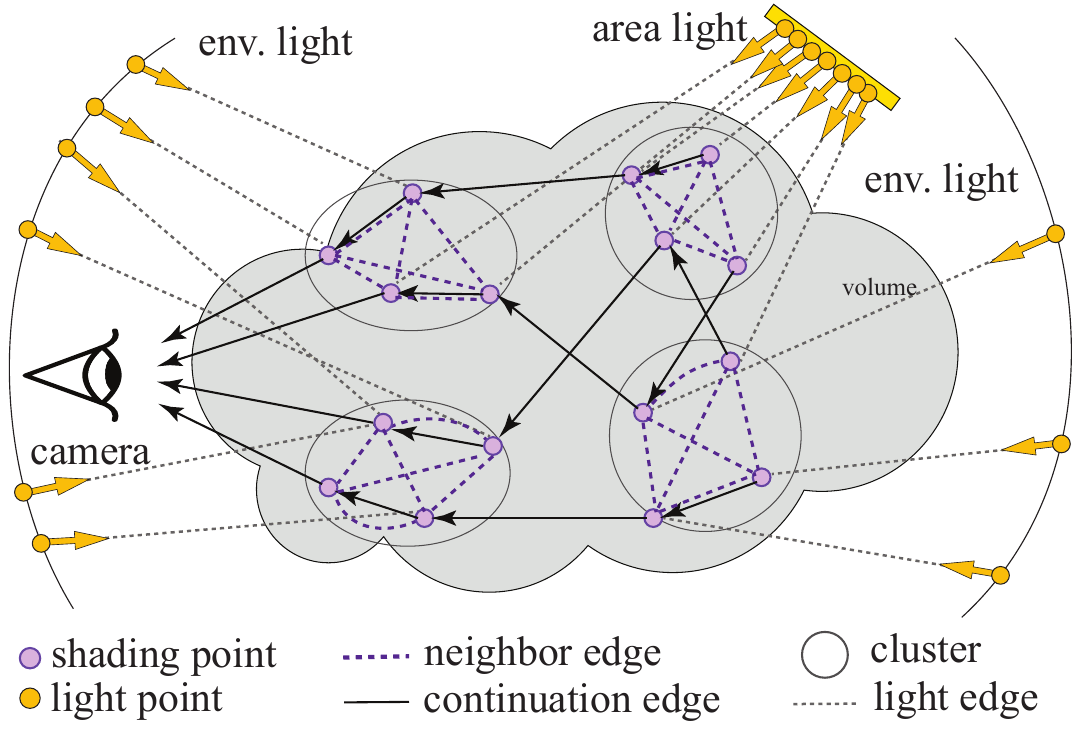}
    \caption{Illustration of the volume path graph construction  from paths sampled during a standard volumetric path tracing pass with next event estimation in participating media.}
    \label{fig:enter-label}
\end{figure}
With \autoref{eq:mc-direct}, \autoref{eq:mc-indirect} and \autoref{eq:mc-pa} one can estimate the integral by taking discrete random samples and continue by recursively expanding the indirect incoming radiance term as another Monte Carlo estimate until the emission is queried. A consecutive chain of shading points $\SPos_0\cdots\SPos_k$ are sampled, with each point representing a scattering event or a surface event.  Typically, each shading point $\SPos_i$ has a connection to an emission point $\LPos_i$ due to next event estimation, completing a light path. When a direction is sampled in proportion to the local phase function, it may also create a connection to light source if there is an unoccluded emission along the direction.
The pixel value is computed from the contribution of those complete light paths. In standard volume path tracing, all intermediate path samples are dropped once the contribution is computed, but in path graph framework, we record enough information to reconstruct these complete light paths and build the information sharing graph. 

\subsection{Aggregation and Propagation}
\label{sec:aggregation and propagation}
After a path graph is constructed, the next step is the aggregation of information  across neighbor edges and propagation along continuation edges. 
We use $i=0\cdots \MaxDepth$ to index shading points on a complete light path $\SPath$ with $ \MaxDepth $ being max depth  of the path, 
and $j=1\cdots K$ to index the shading points within a cluster, with $K$ being the number of shading points in a cluster.

For a shading point $\SPos_i$ in a cluster $\cluster$, we aggregate direct and indirect radiance estimates from all neighbors. One is the incoming indirect radiance $\IndirectIn(\SPos_j, \Dir_j)$, which updates the scattered indirect radiance (noted as $\IndirectOut(\SPos_i, \Dir_i)$) at point $\SPos_i$.
% By aggregation, we assume the incoming radiance distribution is the same for all the point in $\cluster$, instead of sampling $K$ directions on $\SPos_i$ to query $\IndirectIn(\SPos_i, \Dir)$, we directly take incoming radiance estimation $\IndirectIn(\SPos_j, \Dir_j)$ on all $K$ shading points  in $\cluster$ and use them in an approximation of \autoref{eq:mc-indirect}, with optimal combining weights. 
Another quantity we aggregate is the direct radiance $\DirectEmissionIn(\SPos_j, \Dir_j)$, updating refined estimates of scattered direct radiance $\DirectEmissionOut(\SPos_i, \Dir_i)$. Combining those two aggregations gives an updated estimate of total scattered radiance: 
\begin{align}
    \RadianceOut(\SPos_i, \Dir_i) = \IndirectOut(\SPos_i, \Dir_i) + \DirectEmissionOut(\SPos_i, \Dir_i)\label{eq:scatter-sum}
\end{align}
This updated outgoing radiance at $\SPos_i$ corresponds to the incoming indirect radiance at the previous shading point on the path (if any). It is propagated along its outgoing continuation edge to update the indirect radiance $\IndirectIn(\SPos_{i-1}, \Dir_{i-1})$ at the $(i-1)$-th shading point $\SPos_{i-1}$ on $\SPath$.  
In the surface case, \cite{Deng:2021:PathGraph} assumes light transport in vacuum between surfaces, so 
$\IndirectIn(\SPos_{i-1}, \Dir_{i-1}) = \RadianceOut(\SPos_{i},\Dir_i)$.
Unlike the simple copying operation used in the surface case, propagation for the volume path graph should consider the transmittance term and the probability of sampling distance $||\SPos_{i} - \SPos_{i-1}||$ from $\SPos_{i-1}$ in \autoref{eq:mc-pa}, where
\begin{align}
    \IndirectIn(\SPos_{i-1}, \Dir_{i-1}) = \frac{\Transmittance{\SPos_i,\SPos_{i-1}}}{\PdfDist{\SPos_i|\SPos_{i-1}}} \left(\IndirectOut(\SPos_i, \Dir_i) + \DirectEmissionOut(\SPos_i, \Dir_i)\right). \label{eq:volume-propagation}
\end{align}
% \TODO{Maybe I need to use $\SPos_i$ for $\SPos'$, and $\SPos_{i-1}$ for $\SPos$ in \autoref{eq:integral-RTE_3}}

% This aggregation and propagation can repeat and the estimation of $\IndirectOut$ will converge for all the shading points in the path graph. \TODO{mention about the convergence depend on the bsdf weight being less than 1.0 for surface case and this is true for the volume case}
% We will show the proof of convergence of this iteration in the volumetric case later. \TODO{ref to the error analysis section}

% The aggregation process is the only source of bias in path graph system, good thing is that this bias won't result in noticeable artifact nor energy mismatch in the final image. We will show in the next section  (and in the experiments) that this bias is subtle, at the same time, combing the noise reduction due to the aggregation, this error of path graph result is bounded by the noise of path tracing at the same sample rate.

% In the next two subsections we will connect the aggregation and propagation operation with the radiative transfer equation in a participating medium and derive aggregation and propagation operators that are applicable for a path graph built from volume path tracing.

\subsection{Path Graph Operators}
\label{sec:pg-operators}
In path graph, we use an \textbf{aggregation operator} to gather incoming radiance from nearby shading points and recompute the outgoing  radiance at the points. Then we use a \textbf{propagation operator} to propagate the updated quantity to previous shading points on their path. Finally after a few iterations of alternating aggregation and propagation, we propagate updated radiance to the pixels. 

The outgoing radiance of a point is updated by a weighted average of the contribution from nearby shading points in the current cluster. More precisely, this is done by treating other points in the cluster as independent sampling techniques, and combining them by multiple importance sampling. 
Recall that  multiple importance sampling with the balance heuristic \cite{Veach:1995:Optimally} provides weights for combining $m$ sampling strategies as follows:
\begin{align}
    \MISweight(z_i) = \frac{p_i(z_i)}{\sum_{l=1}^{m} p_l(z_i)}, \label{eq:mis-weight}
\end{align}
% \begin{align}
%     \MISweight(z_i) = \frac{n_i p_i(z_i)}{\sum_{l=1}^{m} n_l p_l(z_i)} \label{eq:mis-weight}
% \end{align}
where $p_i(z)$ is the probability of sampling $z$ with the $i$-th strategy. The estimate of $\PContri(z)$ from $m$ strategies is then
% \begin{align}
%     \Est{\PContri(z)} \approx \frac{1}{N} \sum_{j=1}^{N} \MISweight(z_j) \frac{\PContri(z_j)}{p_j(z_j)}. \label{eq:mis-combine-1-sample}
% \end{align}
\begin{align}
    \Est{\PContri(z)} \approx \sum_{j=1}^{m} \MISweight(z_j) \frac{\PContri(z_j)}{p_j(z_j)}. \label{eq:mis-combine-1-sample}
\end{align}
We apply MIS to aggregate the indirect radiance from \autoref{eq:mc-indirect} for point $\SPos$ in a cluster $\cluster$ of size $K$. Given the sampling strategy $\Prior{\SPos_j}$ at point $\SPos_j$ with its sampled direction $\Dir_j$,

\begin{align}
    \IndirectOut(\SPos, \Dir) &=  \sum_{j=1}^{K} \MISweight(\Prior{\SPos}, \Dir_j) \frac{\ScatterFunction{\SPos,\,\Dir,\,\Dir_j} \IndirectIn(\SPos_j, \Dir_j)}{\Pdf{\Prior{\SPos}, \Dir_j}}. \label{eq:aggregate-indirect}
\end{align}
Applying \autoref{eq:mis-weight} to weigh the contribution from direction $\Dir_j$ in cluster $\cluster$ with $K$ points, we get the following weight
\begin{align}
    \MISweight(\Prior{\SPos}, \Dir_j) = \frac{\Pdf{\Prior{\SPos}, \Dir_j}}{\sum_{l=1}^{K} \Pdf{\Prior{\SPos_l}, \Dir_j}}, \label{eq:indirect-weight}
\end{align}
where $\{\SPos_l \in \cluster \}$ and $\Pdf{\Prior{\SPos_l}, \Dir_j}$ is the probability of choosing $\Dir_j$ at $\SPos_l$ using the sampling strategy $\Prior{\SPos_l}$. We call $\PdfM{\Dir_j} = \sum_{l=1}^{K} \Pdf{\Prior{\SPos_l}, \Dir_j}$ the marginal density of directional sample $\Dir_j$ in cluster $\cluster$.
Combining \autoref{eq:indirect-weight} and \autoref{eq:aggregate-indirect} yields:
\begin{align}
     \IndirectOut(\SPos, \Dir) &= \sum_{j=1}^{K} \frac{ \ScatterFunction{\SPos, \Dir, \Dir_j} }{\PdfM{\Dir_j}} \IndirectIn(\SPos_j, \Dir_j).\label{eq:ag-indirect}
\end{align} 
% \XD{need to be clearer about the indexing in the cluster}
The above can be computed for each shading point in time linear in $K$, computing the marginal densities in a first pass and \autoref{eq:ag-indirect} in a second pass over the cluster.

We define $\IndirectOutVec$ as a vector that holds the scattered indirect radiance of all shading points $\SPosUnion$ along all the outgoing continuation edges in $\CEdgeUnion$ and $\IndirectInVec$ as a vector that holds incoming indirect radiance at all the shading points $\SPosUnion$ from all the incoming continuation edges in $\CEdgeUnion$. Then \autoref{eq:ag-indirect} can be expressed in matrix form as follow
\begin{align}
    \IndirectOutVec = \AiMatrix\IndirectInVec,
    \label{eq:ag-indireect-mf}
\end{align}
where
$\AiMatrix$ is a $|\SPosUnion|\times|\SPosUnion|$ matrix with its element on row $r$ column $c$ being
\begin{align}
    \AiMatrix_{rc} = \begin{cases} \frac{\ScatterFunction{\SPos_r, \Dir_r, \Dir_c}}{\PdfM{\Dir_c}}\,, &\SPos_r, \SPos_c \in \text{the same cluster} \\
        0\,, & \SPos_r, \SPos_c \, \text{in different clusters}.
    \end{cases}
\end{align}

% then we write \autoref{eq:ag-indirect} in the following matrix from

Aggregation of the direct radiance is slightly different from the indirect aggregation. For each shading point $\SPos$ in the cluster, it has two directions sampled for the incoming direct radiance, one from emitters and one from the local phase function. Therefore, when combining the estimations from all the direct radiance samples in a cluster, we consider $2K$ samples from $K+1$ sampling strategies: $K$ samples from emitter sampling (here we view all emitter sampling as one strategy), and $K$ local phase function strategies for each sample. %By MISing with multi-sample for each strategy, \autoref{eq:mis-weight} and \autoref{eq:mis-combine-1-sample} change to 
% \begin{align}
%     \Est{\PContri(z)} \approx \sum_{i=1}^{m} \frac{1}{n_i} \sum_{j=1}^{n_i} \MISweight(z_j) \frac{\PContri(z_j)}{p_j(z_j)}, && \MISweight(z_i) = \frac{n_i p_i(z_i)}{\sum_{l=1}^{m} n_l p_l(z_i)}, \label{eq:mis-n-sample}
% \end{align}
% \begin{align}
%     \MISweight(z_i) = \frac{n_i p_i(z_i)}{\sum_{l=1}^{m} n_l p_l(z_i)} \label{eq:mis-weight}
% \end{align}
%where $n_i$ is the number of samples from the $i$-th strategy.
Applying this multi-sample MIS on the $2K$ direct light samples for \autoref{eq:mc-direct} and cancelling some terms gives us
\begin{align}
\DirectEmissionOut(\SPos,\Dir) &=  \frac{1}{K} \sum_{j=1}^{K} \frac{K \ScatterFunction{\SPos, \Dir, \DirE_j} \DirectEmissionIn(\SPos_j, \DirE_j)}{\sum_{l=1}^{K} \PdfP{\Prior{\SPos_l}, \DirE_j} + K \PdfE{\DirE_j}} \\&+   \sum_{j=1}^{K} \frac{ \ScatterFunction{\SPos, \Dir, \Dir_j} \DirectEmissionIn(\SPos_j, \Dir_j)}{\sum_{l=1}^{K} \PdfP{\Prior{\SPos_l}, \Dir_j} + K \PdfE{\Dir_j}},
\end{align}
where $\DirE$ stands for the directional sample taken from emitter sampling with a PDF $\PdfE{\DirE}$, and $\PdfP{\Prior{\SPos_l}, \Dir}$ is the PDF of sampling $\Dir$ from phase function at a shading point $\SPos_l$.
Similar to the indirect radiance case, we define $\PdfM{\Dir} = \sum_{l=1}^{K} \PdfP{\Prior{\SPos_l}, \Dir} + K \PdfE{\Dir_j}$ as the marginal density for direction $\Dir$ in its cluster, which gives us
\begin{align}
    \DirectEmissionOut(\SPos,\Dir) &= \sum_{j=1}^{K} \frac{ \ScatterFunction{\SPos, \Dir, \DirE_j} }{\PdfM{\DirE_j}} \DirectEmissionIn(\SPos_j, \DirE_j) +  \sum_{j=1}^{K} \frac{ \ScatterFunction{\SPos, \Dir, \Dir_j}}{\PdfM{\Dir_j}} \DirectEmissionIn(\SPos_j, \Dir_j).\label{eq:ag-direct}
\end{align}
To turn \autoref{eq:ag-direct} into a matrix form, we define $\DirectEmissionInVec$ as a vector of length $|\LPosUnion| = 2|\SPosUnion|$, with interleaved entries being incoming direct radiance from the light edges on the path graph which are sampled from the emitter distribution (entries with even indices) or in proportion to the local phase function (entries with odd indices); then $\DirectEmissionOutVec$ is a vector of size $|\SPosUnion|$ holding the scattered direct radiance on all the shading points along all continuation edges. Writing \autoref{eq:ag-direct} in a matrix form, we have
\begin{align}
    \DirectEmissionOutVec =  \AdMatrix \DirectEmissionInVec,\label{eq:ag-direect-mf}
\end{align}
where the direct aggregation matrix $\AdMatrix$ has a size of $|\SPosUnion|\times|\LPosUnion|$;
% \TODO{move those details to supplementary}
its entry on row $r$ and column $c$ is
\begin{align}
    \AdMatrix_{rc} = \begin{cases}
        \frac{\ScatterFunction{\SPos_r, \Dir_r, \DirE_{c/2}}}{\PdfM{\DirE_{c/2}}}\,, & \SPos_r, \SPos_{c/2} \, \text{in same cluster and c is even} \\
        \frac{\ScatterFunction{\SPos_r, \Dir_r, \Dir_{c/2}}}{\PdfM{\Dir_{c/2}}}\,, & \SPos_r, \SPos_{c/2} \,\text{in same cluster and c is odd} \\
        0, & \text{otherwise}.
    \end{cases}
\end{align}
% Then \autoref{eq:ag-direct} can be write as 
% \STEVE{We need to explain the indexing of $L^o$, where it has length $2|V_y|$ with interleaved entries}

Putting \autoref{eq:ag-direect-mf} and \autoref{eq:ag-indireect-mf} into \autoref{eq:scatter-sum} we get 
\begin{align}
    \RadianceOutVec = \IndirectOutVec + \DirectEmissionOutVec = \AiMatrix\IndirectInVec + \AdMatrix \DirectEmissionInVec \label{eq:ag-mf}.
\end{align}
\autoref{eq:ag-mf} is the aggregation operation; it gives us the equations for recomputing the outgoing radiance at each shading point from the recorded quantities on the path graph.

Once the outgoing radiance $\RadianceOutVec$ of a shading point is refined, we need to propagate this refinement towards the previous shading point through the continuations edges.  This propagation operation is described in \autoref{eq:volume-propagation}, and we also write it in matrix multiplication form as %on path graph
\begin{align}
    \IndirectInVec = \PMatrix \RadianceOutVec\label{eq:rte-pg}
\end{align}
with a propagation matrix $\PMatrix$ of size $|\SPosUnion|^2$ whose entry at row $r$ and column $c$ is
\begin{align}
    \PMatrix_{rc} = \begin{cases}
        \frac{\Transmittance{\SPos_r,\SPos_c}}{\PdfDist{\SPos_r|\SPos_c}}\,, & \SPos_r \text{is the previous shading point of }\SPos_c \\
        0.0\,, & \text{otherwise}.
    \end{cases}
\end{align}
Expanding \autoref{eq:rte-pg} by \autoref{eq:ag-mf} yields
\begin{align}
    \IndirectInVec = \PMatrix\AiMatrix\IndirectInVec + \PMatrix\AdMatrix\DirectEmissionInVec,\label{eq:rte-mff}
\end{align}

Now that we have introduced the aggregation operators ($\AdMatrix, \AiMatrix$) and the propagation operator ($\PMatrix$) over the path graph, given a path graph $\PathGraph = \langle \VertexUnion, \EdgeUnion \rangle$ and the recorded variables on the path graph, we are solving for a fixed point $\IndirectInVec$ for the updating rule in \autoref{eq:rte-mff}, and each image pixel value can be reconstructed by selecting the entries of $\IndirectInVec$ that tie to that pixel. 
Since the direct illumination vector $\PMatrix\AdMatrix\DirectEmissionIn$ is independent of the variable $\IndirectInVec$ that is being updated, it is computed once and remains constant throughout the iteration. With $\IndirectInVec$ initialized to the indirect radiance estimates from path tracing, the linear system converges within a few iterations.

\section{Rendering System}
\label{sec:rendering-system}
We implemented an instrumented volumetric path tracer for path graph recording based on the open source renderer Mitsuba v0.6 \cite{Jakob:2013:Mitsuba} and added our volumetric path graph solver as a CUDA plugin. 

% instrumented path tracing: handling volume boundary;
\Paragraph{Instrumented path tracer} Path tracing is performed on the CPU. During path tracing, we record data for the emitter sampling and phase function sampling at each shading point. Additionally, the incoming indirect radiance for each shading point is recorded. Once a path is traced, all of its shading points are stored into a global array $\SPos$ that preserves their order on the path. %(i.e. $\SPos_i$ is the previous shading point (counting from the camera) of shading point $\SPos_{i+1}$, unless $\SPos_{i+1}$ is the first point on its path, in which case $\SPos_i$ is the last point on another path).
$\SPos$ is stored in pinned memory to minimize data transfer time onto the GPU.

\Paragraph{Path graph solver} Once the shading points are on the GPU, $m$ cluster centers are randomly selected such that $mK=n$ where $K$ is the expected cluster size and $n$ is the total number of shading points. A hash-grid on the shading points' positions is built, and a nearest-neighbor clustering is performed by searching through neighboring bins in the hash grid. After forming the clusters, we use data from the recorded light points to perform one pass of direct radiance aggregation following \autoref{eq:ag-direct}. We also use the recorded incoming indirect radiance for each shading point to compute the aggregated indirect radiance by following \autoref{eq:ag-indirect}. The sum of aggregated direct and indirect light is then propagated across each continuation edge. We then perform a few iterations of alternating indirect radiance aggregation and propagation to get to the final result. Note that we could have initialized a zero-valued incoming indirect radiance at each shading point, but it would take $q$ iterations of aggregation and propagation to correctly compute the radiance for a path of length $q$; that is not ideal for volumetric rendering, which can have especially long paths. Starting from path-tracing estimates of indirect incoming radiance puts the image at the correct energy level from the first iteration, and the iterative refinements are only used to improve the image smoothness. It normally takes a small number of iterations (less than 10) to obtain a converged output, even if many paths are much longer than 10 bounces.

\Paragraph{Handling direct Light} Direct radiance aggregation is only performed to better optimize indirect radiance (through propagation). In order to remove significant correlations between neighboring points, we  use the non-aggregated direct radiance from path tracing instead of the aggregated direct radiance when writing the final output. This means that our method can be expected mainly to improve on scenes dominated by indirect radiance. However, for certain scenes that are dominated by participating media but have a few direct radiance components (e.g. emitters in a medium or media encapsulated by dielectric surfaces), Path Graph can still show a benefit, but only if the direct radiance is handled well. To this end we optionally record a few extra emitter samples for the first bounce on each path. In scenes where direct radiance needs more improvement, we apply Path Graph as normal by using the aggregated direct radiance in propagation, but then use the extra direct light samples, instead of the non-aggregated direct radiance, to compute the final pixel values.

\Paragraph{Final output} The radiance on the continuation edge for the first shading point of each path represents the radiance propagated into the camera. These radiance values will be the output for the Path Graph after a few iterations of aggregation and propagation. The output values for each path are written into the corresponding pixels of the output image, which can then be optionally fed into a denoiser for the final output.

\section{Experiments}
\begin{table*}[]
\caption{Detailed timing of our pipeline steps. The 2nd to 5th columns are the average render time of 1 sample per pixel for both path tracing, revised path tracing with data recording, recording of extra direct light samples (only applicable to \buddha \ and \trafficlight), and the total execution time of our path graph iteration. The ratio is the number of samples that a path tracer could compute during the same time of 1spp using our method, with path sample recording overhead. The number of PT/PG samples used in 5 min equal time comparison is computed from the time for a single sample run, with an additional note that if extra direct light samples are recorded, this operation will only be performed once and thus its time would not multiply with the number of Path Graph samples.}

\resizebox{\textwidth}{!}{%
\begin{tabular}{|c|c|c|c|c|c|c|c|c|c|}
\hline
\multirow{3}{*}{Scene Name} & \multicolumn{4}{c|}{time ( s )} & \multirow{3}{*}{ratio} & \multicolumn{2}{c|}{\begin{tabular}[c]{@{}c@{}}MSE\\ for 1SPP\end{tabular}} & \multirow{3}{*}{\begin{tabular}[c]{@{}c@{}}pathtr:ours SPP\\ for equal time(300s)\end{tabular}} & \multirow{3}{*}{\begin{tabular}[c]{@{}c@{}}Equal time (300s) MSE ratio \\ path tr: ours\end{tabular}} \\
\cline{2-5} \cline{7-8}
 & \multirow{2}{*}{path tr} & \multicolumn{3}{c|}{ours} & &  \multirow{2}{*}{\begin{tabular}[c]{@{}c@{}}path \\ tr\end{tabular}} & \multirow{2}{*}{ours} & & \\ \cline{3-5}
 &  & record path tr & record extra direct samples & path graph &  & \multicolumn{1}{l|}{} &  &  &   \\ \hline
{\sc Buddha} & 18.9 & \REMOVE{1.2} \ADD{24.2} & 2.8 (15spp) & 16.92 & \REMOVE{3.7} \ADD{2.32} & 0.0184 & 0.000837 & 15:\ADD{7} & \ADD{1.24e-3 : 2.68e-4 = 4.62}\\ \hline
{\sc Traffic-Light} & 4.3 & \REMOVE{1.2} \ADD{6.0} & 31.1 (69spp) & 9.426 & \REMOVE{3.7} \ADD{10.82} & 0.0263 & 0.000999 & 69:\ADD{17} & \ADD{8.98e-4 : 2.12e-4 = 4.24}\\ \hline
{\sc Foggy-Forest} & 6.8 & \REMOVE{1.2} \ADD{10.4} & N/A & 13.74 & \REMOVE{3.7} \ADD{3.55} & 0.0708 & 0.00515 & 44:\ADD{12} & \ADD{1.61e-3 : 4.34e-4 = 3.70}\\ \hline
{\sc Disney-Cloud} & 3.0 & \REMOVE{1.3} \ADD{15.8} & N/A & 5.148 & \REMOVE{7.5} \ADD{6.98} & 0.0786 & 0.00391 & 100:\ADD{14} & \ADD{7.87e-4 : 3.07e-4 = 2.57} \\ \hline
% {\sc Dielectric-Dragon} & 8.8 & \REMOVE{1.2} \ADD{10.5} & 3.916+0.561+13.581+2.265 (=20.32) & \REMOVE{3.1} \ADD{3.50} &  &  & 13:\ADD{3,34:9} & \ADD{0.000356/0.000682=0.52} \\ \hline
{\sc Industry-Smoke} & 12.5 & \REMOVE{1.3} \ADD{13.1} & N/A & 1.826 & \REMOVE{2.4} \ADD{1.25} & 0.0432 & 0.00721 & 24:\ADD{19} & \ADD{1.80e-3 : 3.84e-4 = 4.69} \\ \hline
% {\sc AerialExplosion} & 6.0 & \REMOVE{1.2} \ADD{6.7} & 1.621+0.128+0.236+0.992 (=2.619) & \REMOVE{3.6} \ADD{1.55} &  &  & 20:\ADD{12,50:32} & \ADD{34.28/9.44=3.63} \\ \hline
% {\sc tornado} & 23.7 & \REMOVE{36.7} \ADD{36.7} & 2.18+0.196+0.407+1.13 (=3.91) & \REMOVE{4.0} \ADD{1.71} &  &  & 5:\ADD{2,12:7} & \ADD{0.000238/0.000469=0.51} \\ \hline
{\sc Mitsuba-Colored-Smoke} & 6.2 & \REMOVE{1.2} \ADD{6.9} & N/A & 6.882 & \REMOVE{3.6} \ADD{2.22} & 0.212 & 0.0238 & 48:\ADD{21} & \ADD{4.31e-3 : 3.95e-4 = 10.91} \\ \hline
{\sc Golden-Gate} & 43.8 & \REMOVE{1.2} \ADD{50.4} & N/A & 17.36 & \REMOVE{3.7} \ADD{1.55} & 0.128 & 0.0135 & 7:\ADD{4} & \ADD{1.85e-2 : 3.48e-3 = 5.32}\\ \hline
% {\sc DuskShockWave} & 9.5 & \REMOVE{1.1} \ADD{13.2} & 1.577+0.12+0.196+0.26+0.25(=2.40) & \REMOVE{2.4} \ADD{1.64} &  &  & 12:\ADD{7,31:19} & \ADD{}\\ \hline
\end{tabular}%
}
\label{tab:my-table}
\end{table*}
We study the behavior of path graph on a variety of volumetric scattering scenes under challenging lighting conditions. Our tests contain both homogeneous (\buddha, \trafficlight, \foggyforest) and heterogeneous volumes (\bunnycloud, \disneycloud, \dustshockwave, \mtscolorsmoke, \industrysmoke, and \goldengate). The resolution of all renderings is 1440x960 pixels.
The CPU we use is an Intel Xeon Silver 4214 with 8 cores, and the GPU is an Nvidia RTX 3090.

\begin{figure*}
% [t!]
    \centering
    \setlength{\fboxsep}{0pt}%
    \setlength{\fboxrule}{1pt}%
    \contourlength{0.1em}%
    \setlength{\mygrid}{1\textwidth}%
    \setlength{\myimg}{0.99\textwidth}%
    \hspace*{-3ex}%
    \begin{tikzpicture}[x=\mygrid,y=0.5\mygrid,every text node part/.style={align=center}]
        % \node at (2,    0.72) {$(\Ua,t)$-planes};
        % \node at (2.5,  0.72) {$+$};
        % \node at (3,    0.72) {$(u,v)$-planes};
        % \node at (3.5,  0.72) {$=$};
        % \node at (4,    0.72) {Combination};
        % \node at (5+.05, 0.72) {Beams};
        % \node at (2+.25,.58) {MIS};
        % \node at (2-.25,.58) {AVG};
        \node at (0,  0) {\includegraphics[width=\myimg]{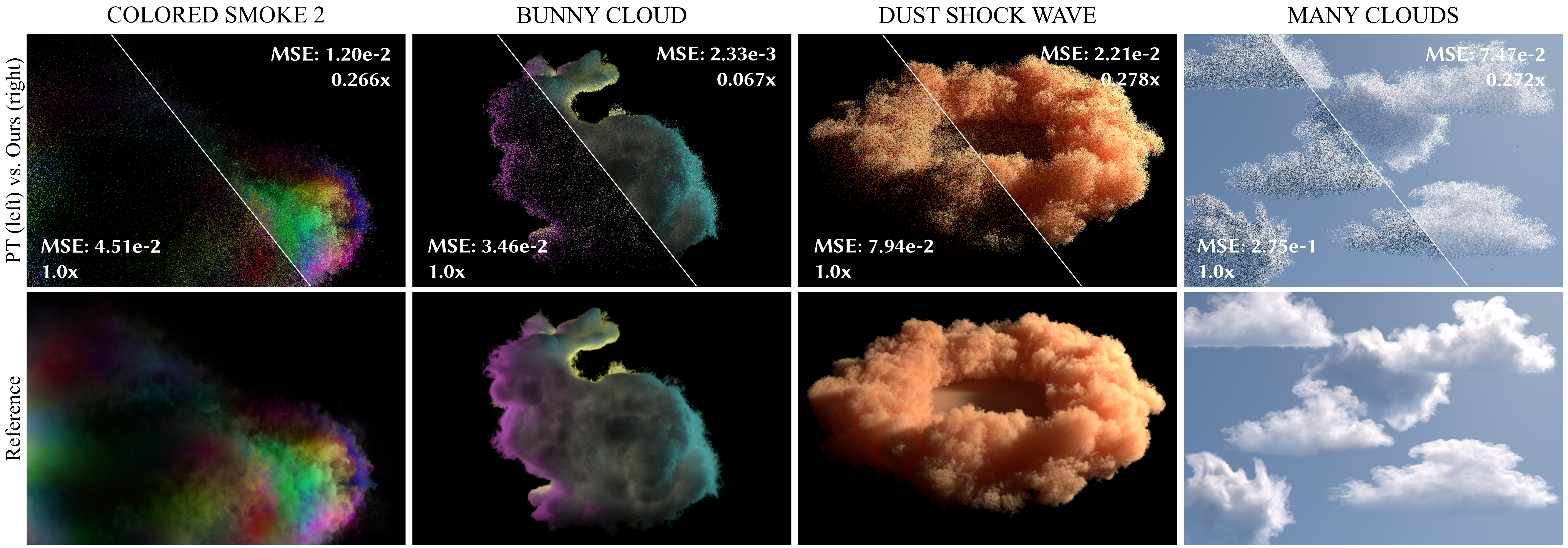}};
    \end{tikzpicture}%
    \vspace*{-1\baselineskip}%
    \caption{
    \ADD{We show an \textbf{equal sample (at 1 sample per pixel) comparison} between ours and path tracing. Our method extensively reuses the paths' information even with only one sample per pixel, significantly improving rendering efficiency.}
    }
    \label{fig:1spp-comparison-hetero}%
    % \vspace*{-1.5\baselineskip}%
\end{figure*}

\Paragraph{Equal sample comparison}
In \autoref{fig:1spp-comparison-hetero} we show an 1 spp comparison between our method and path tracing on scenes with heterogeneous media. Our method is much closer to the reference and produces a much smaller MSE than path tracing.

\begin{figure*}
% [t!]
    \centering
    \setlength{\fboxsep}{0pt}%
    \setlength{\fboxrule}{1pt}%
    \contourlength{0.1em}%
    \setlength{\mygrid}{0.2\textwidth}%
    \setlength{\myimg}{0.199\textwidth}%
    \hspace*{-3ex}%
    \begin{tikzpicture}[x=\mygrid,y=\mygrid,every text node part/.style={align=center}]
        \node at (1,  0) {\includegraphics[width=\textwidth]{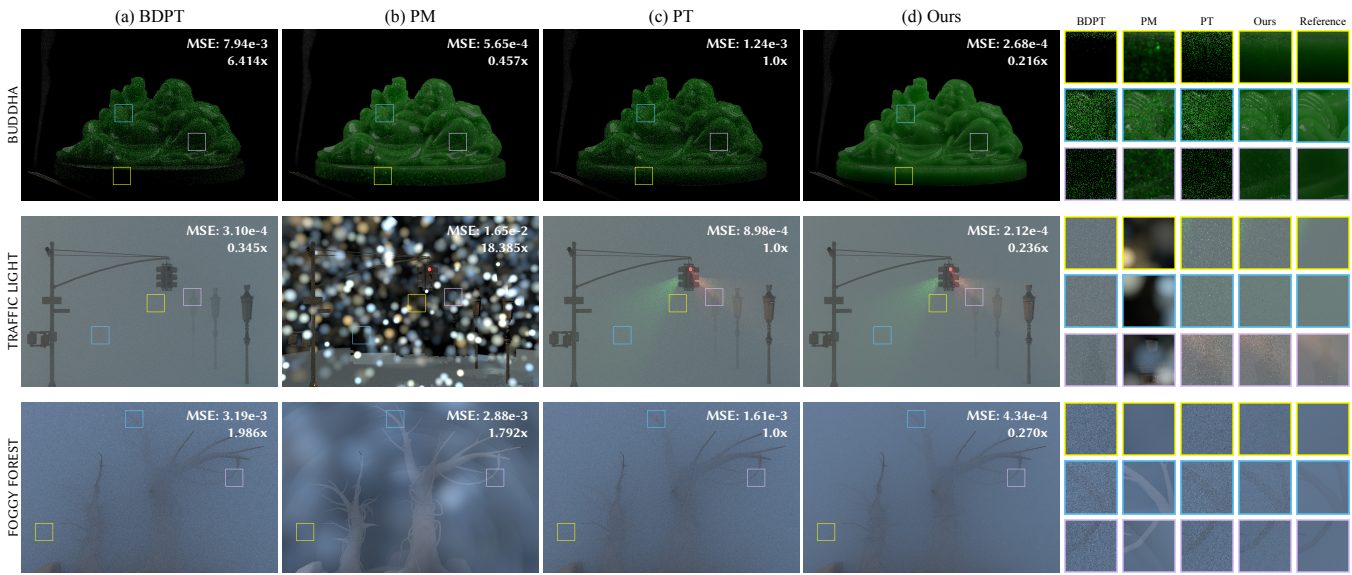}};
    \end{tikzpicture}%
    % \vspace*{-1\baselineskip}%
    \caption{
    \ADD{Comparison between (a) bidirectional path tracing, (b) photon mapping, (c) path tracing and (d) our method (path tracing + path graphs) on an \textbf{equal time (5 min)} in scenes with the presence of homogeneous volume. We show full light transport in all the scenes and our method provides significant variance reduction over previous methods on rendering the participating media. The $2^{\text{nd}}$ and $3^{\text{rd}}$ scenes contain large volume of outdoor medium (e.g. fog), where (unguided) photon mapping suffers since only a small amount of photons arrive at the region where the camera is looking towards due to strong multiple scattering. The $2^{\text{nd}}$ scene has a skydom as environment lighting as well as the traffic lights, where the bidirectional method failed to sample the small light sources.}
    }
    \label{fig:homogeneous-equal-time}%
    % \vspace*{-1.5\baselineskip}%
\end{figure*}

\begin{figure*}
% [t!]
    \centering
    \setlength{\fboxsep}{0pt}%
    \setlength{\fboxrule}{1pt}%
    \contourlength{0.1em}%
    \setlength{\mygrid}{0.166\textwidth}%
    \setlength{\myimg}{0.16\textwidth}%
    \hspace*{-3ex}%
    \begin{tikzpicture}[x=\mygrid,y=\mygrid,every text node part/.style={align=center},spy using outlines={width=28mm,height=6mm}]
        \node at (1.0,  0) {\includegraphics[width=\textwidth]{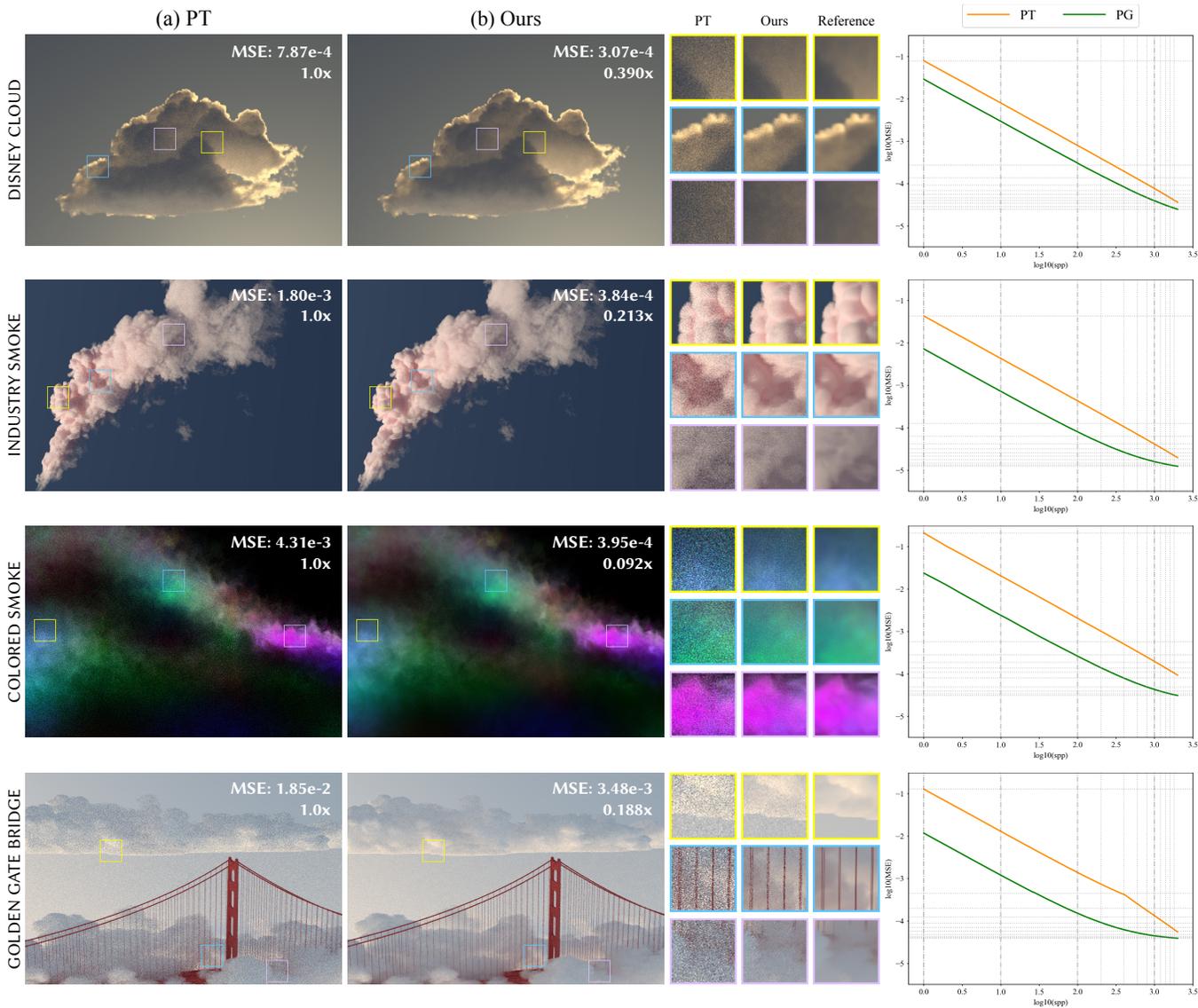}};

    \end{tikzpicture}%
    \caption{
    \ADD{Comparison between (a) path tracing and (b) our method (path tracing + path graphs) on an \textbf{equal time (5 min)} in scenes with the presence of heterogeneous volume. We show full light transport in all the image and our method out perform the path tracing especially around the area where the pixel value is dominated by multi-bounces.}
    }
    \label{fig:heterogeneous-equal-time}%
\end{figure*}

\Paragraph{Equal time comparison}
Our method necessarily has some overhead per sample. Therefore, we also render some more challenging scenes using both path tracing and our method in 5 minutes. In all tested cases, our method still visually gives a smoother result than path tracing and numerically has a smaller MSE, as we can see in \autoref{fig:homogeneous-equal-time} and \autoref{fig:heterogeneous-equal-time}. The detailed breakdown of path tracing and our method's time and MSE is shown in \autoref{tab:my-table}. Our method produces even more variance reduction in scenes that have more indirect light (e.g. \mtscolorsmoke) and/or media with a high density and high albedo (e.g. \industrysmoke).

Note that despite being dominated by indirect light, \buddha \ and \trafficlight \ both have some noticeable direct light components (specular reflections on the dielectric surface in \buddha \ and the emitters in \trafficlight), and as mentioned in \autoref{sec:rendering-system}, here we used additional direct light samples to render the final result. Again these 1-bounce samples are not used in the path graph itself but only  when writing the final output to make up for the path graph's non-aggregated direct radiance. We make it so that path graph has an equal number of direct illumination samples as path tracing, and thus any variance reduction comes from path graph itself. The time for the additional samples is accounted for in the equal time comparison. Since the Path Graph already handles the more difficult indirect illumination, the direct illumination samples do not take long to render and won't significantly reduce the number of path graph samples that could be run. We can see that path graph still gives more desirable results in both scenes. 

\Paragraph{Photon mapping and BDPT}
Photon mapping and Bidirectional Path Tracing (BDPT) are two well-known methods for rendering homogeneous media. We include them in our equal time comparison shown in \autoref{fig:homogeneous-equal-time}. These two methods show certain strengths in some types of scenes but both have significant limitations. Photon mapping gives a reasonable result in a small enclosed medium (\buddha), but it struggles particularly in outdoor scenes (\trafficlight \ and \foggyforest) where targeting the photons into the camera view is non-trivial. BDPT, though theoretically unbiased, converges slowly and thus has a larger MSE when rendered in a small amount of time like here. Our method adapts well to the various lighting conditions and has barely noticeable bias.

\Paragraph{Log-log error curves}
In \autoref{fig:heterogeneous-equal-time}, we further show a log-log convergence curve up to 2000 iterations of path tracing and our method. Due to the bias we introduce during aggregation, the curve for our method will eventually intersect with path tracing's curve when it is more preferable to use path tracing. Nevertheless, this will happen well beyond 2000 spp, showing that our method can improve convergence across a reasonable range of spp settings.

\begin{figure*}
% [t!]
    \centering
    \setlength{\fboxsep}{0pt}%
    \setlength{\fboxrule}{1pt}%
    \contourlength{0.1em}%
    \setlength{\mygrid}{1\textwidth}%
    \setlength{\myimg}{0.99\textwidth}%
    \hspace*{-3ex}%
    \begin{tikzpicture}[x=\mygrid,y=0.5\mygrid,every text node part/.style={align=center}]
        \node at (0,  0) {\includegraphics[width=\myimg]{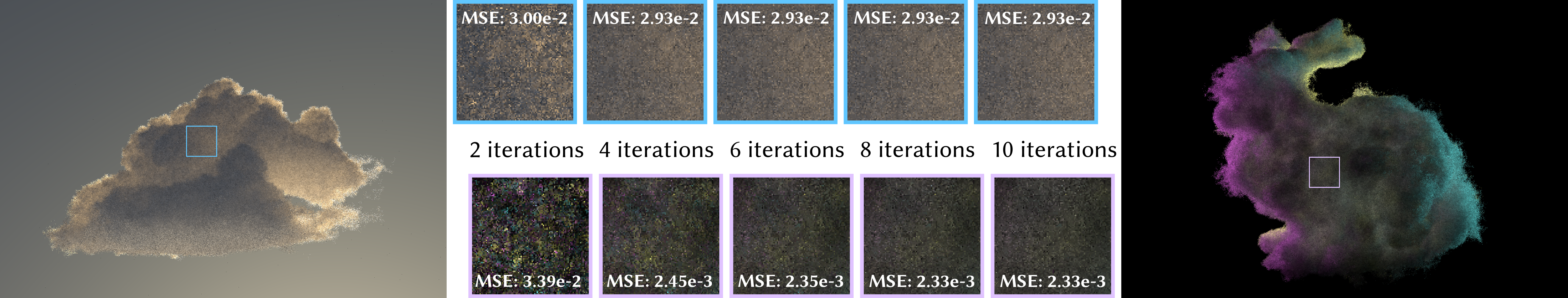}};
    \end{tikzpicture}%
    \vspace*{-1\baselineskip}%
    \caption{
    \ADD{Two 1spp scenes with a varying number of Path Graph iterations. Path Graph converges well within 10 iterations in both cases.}
    }
    \label{fig:PG-iterations}%
    % \vspace*{-1.5\baselineskip}%
\end{figure*}

\Paragraph{Study of Path Graph iterations}
As mentioned in \autoref{sec:rendering-system}, it often takes less than 10 iterations of aggregation and propagation for Path Graph to converge. To support our claim, we show in \autoref{fig:PG-iterations} how Path Graph's output changes as the number of iterations varies. It can be seen that even though the two scenes differ in the amount of direct illumination (the bunny scene has more indirect light and thus takes a few more iterations to converge), they both converge within 10 iterations.

% \Paragraph{Study of the cluster size}

% \section{Discussion}
% TBD. 

\section{Conclusion}
In this work, we extended the path graph framework to scenes involving volumetric scattering media like clouds and fog. Our method focuses on more effective reuse of multiple-scattering paths and offers a notable improvement in sampling efficiency and speed over traditional path tracing techniques, particularly in challenging environments with heterogeneous, highly scattering media and complex lighting. While our approach shows significant potential in enhancing volumetric rendering, it represents just one step in addressing the broader challenges of the field. Future explorations could include deeper integration of volume and surface path graphs.

% equal time heterogeneous

%% Acknowledgements
% \begin{acks}
% Acknowledgements
% \end{acks}

%% Bibliography.
%% Uncomment the bibliography line and link to an actual bib file
\bibliographystyle{ACM-Reference-Format}
\bibliography{rendering-bibtex}

%% Appendix
% \appendix

% \section{Additional Section}

% Text

\end{document}